\documentclass[sigconf, screen]{acmart}
 
\AtBeginDocument{%
  \providecommand\BibTeX{{%
    \normalfont B\kern-.05em{\sc i\kern-.025em b}\kern-.08em
    T\kern-.1667em\lower.7ex\hbox{E}\kern-.125emX}}}
 
\setcopyright{none}
\acmConference[GI 2026]{Graphics Interface 2026}{June 9--12, 2026}{Kitchener-Waterloo, ON, Canada}
\acmYear{2026}
\renewcommand\footnotetextcopyrightpermission[1]{}

\begin{document}
 
\title{Interpretive Proactivity: Expanding the Design Space of Proactive Interaction}

\author{Pengqi Wang}
\affiliation{%
  \institution{York University}
  \city{Toronto}
  \country{Canada}}
\email{ericwang@yorku.ca}
 
\author{Emily Kuang}
\affiliation{%
  \institution{York University}
  \city{Toronto}
  \country{Canada}}
\email{ekuang@yorku.ca}
 
\renewcommand{\shortauthors}{Wang and Kuang}
 
\begin{abstract}
Research on proactive systems in HCI has grown substantially, but the discussion has centered on systems that proactively generate actions, recommendations, and anticipatory responses for users.
We identify a second paradigm of proactive behavior that is just as widespread and far less examined. Input method editors, auto-correction, and similar text-entry systems take initiative on every keystroke, not to generate new content but to infer, interpret, and commit to meanings from partial user input.
We call the first paradigm \textit{generative proactivity} and the second \textit{interpretive proactivity}, and we argue that this second paradigm deserves a place of its own in the proactivity research agenda. 
Using multilingual input and auto-correction as case studies, we show that interpretive proactivity inherits the vocabulary of proactive design but changes what that vocabulary refers to. Timing becomes a question of when to commit an interpretation rather than when to interrupt, and control calls for the ability to reinterpret committed input rather than only accept or reject a suggestion. 
We close by arguing that the cost of errors should drive the design of interpretive systems, and posing open questions for the workshop.
\end{abstract}
 
\maketitle

\section{Introduction}

The HCI community's discussion of proactive systems has grown rapidly in the last few years.
Voice assistants that anticipate user needs, conversational agents that take initiative in dialogue, 
coding copilots that suggest the next line of code, all of which are routinely described as proactive systems~\cite{dengHumancenteredProactive2024,kuangEnhancingUX2024}. 
The PROACT workshop's call points out the problem that this growth has created: the word \emph{proactive} now covers very different system behaviors, from simple reminders to goal-anticipating assistants, and this conceptual ambiguity makes it hard to design and evaluate these systems with rigor.

In this position paper, we argue that the current conversation around proactivity has settled around a particular %
paradigm without fully recognizing it.
The proactive systems most often studied share %
an underlying logic.
They observe the user's context, decide that some new content or action would help, and surface it as something the user can accept, reject, or refine. Recommendations, suggested replies, scheduled appointments, conversational openings, and writing completions all follow this paradigm~\cite{horvitzPrinciplesMixedinitiative1999,shiHCICentricSurvey2024}. 
We call this \emph{generative proactivity}, because the system's contribution is something the user did not yet have.

A second paradigm of proactive behavior sits in plain sight but rarely enters the conversation. 
Input method editors (IMEs), autocorrect, swipe keyboards, context-sensitive toolbars, and spelling correction all take initiative on every user action, without ever being asked. 
What they produce is an interpretation of the input the user has already provided. The system decides what the keystroke stream means, commits to that interpretation, and lets the user keep typing. 
We call this \emph{interpretive proactivity}.  These systems take initiative under uncertainty, which is the core concept of proactivity in Horvitz's classic framing~\cite{horvitzPrinciplesMixedinitiative1999}, yet the proactivity discussion has grown around content production and turn-based suggestion timing. For instance, Kristensson's survey of intelligent text entry, the closest existing work, does not use the word \emph{proactive} anywhere~\cite{kristenssonFiveChallenges2009}.

However, the two paradigms are not mutually exclusive. Many systems can do both, and interpretation often runs underneath generation. For example, a tool that suggests a whole phrase must first interpret what has been typed. We believe that the interpretive mode, despite being pervasive, has gone underexamined on its own terms.
Even so, the distinction matters because the two paradigms face different design pressures. 
Generative proactivity runs at the pace of user reflection, where seconds of latency are acceptable, and a wrong suggestion is usually just dismissed. Interpretive proactivity runs at the pace of motor action, where milliseconds matter and a wrong interpretation becomes text that the user has to undo against the flow of writing. Frameworks built for the first paradigm carry over awkwardly. 
Suggestion-timing studies in the generative case~\cite{kuangEnhancingUX2024} do not map onto the question of when to commit to a sub-second interpretation. 
Explanation-based transparency is hard to apply when a system makes a decision on every keystroke and cannot pause to justify each one. Moreover, control mechanisms built around accepting or rejecting a discrete suggestion do not help much once the system has already interpreted a whole phrase the wrong way.

This paper makes three contributions to the workshop's goals. 
First, we name the generative and interpretive distinction as two paradigms of proactive design and argue that interpretive proactivity is an appropriate and underdeveloped part of the proactivity research agenda. 
Second, we use multilingual input and auto-correction as cases, drawing on an interactive prototype we have built that varies systems of this kind along several design dimensions. 
Third, we work through what interpretive proactivity forces us to reconsider the four dimensions the workshop has identified as central: timing, transparency, control, and the cost of errors. We close with open questions we hope to discuss at the workshop.

\section{Two Paradigms of Proactivity}

The word \emph{proactive} has a long history in computing. Tennenhouse's 2000 manifesto used it to describe a move beyond human-centered interaction, with systems that sense, decide, and act on the world at faster-than-human time scales~\cite{tennenhouseProactiveComputing2000}. His three pillars (getting physical, getting real, getting humans out of the loop) all point toward systems that initiate action in the environment. A year earlier, Horvitz had given the field a decision-theoretic vocabulary for when an agent should act, ask, or wait, based on expected utility under uncertainty about user goals~\cite{horvitzPrinciplesMixedinitiative1999}. The modern lineage runs through proactive conversational agents, %
where \citet{dengHumancenteredProactive2024}  organized the design space around intelligence, adaptivity, and civility, and through broader human-AI collaboration work, where \citet{shiHCICentricSurvey2024} scoped a recent survey explicitly to generative AI.

Across this lineage, the standard proactive system shares recognizable logics. It keeps a probabilistic model of the user's goal or situation, decides at some moment that a useful action or piece of content can be offered, and surfaces that contribution for the user to evaluate.  Horvitz's LookOut infers a scheduling intent from email content and offers to create an appointment~\cite{horvitzPrinciplesMixedinitiative1999}. Smart Reply guesses that a short response is wanted and proposes three options. Copilot guesses the next line of code.  \citet{kuangEnhancingUX2024} studied this paradigm directly in UX evaluation, asking whether ChatGPT-generated problem suggestions should appear before, during, or after the relevant moment in a usability test video, and found that evaluators strongly preferred suggestions delivered after they had formed their own interpretation. The common feature is that the system's contribution is a new artifact, delivered at a discrete moment, and the user's response is to judge it.

A different paradigm of proactive system has been studied separately, under the heading of intelligent text entry.
Kristensson and Jameson's survey of intelligent text entry treats these as systems that combine AI and HCI to reduce the user's articulation effort~\cite{kristenssonFiveChallenges2009}. Each keystroke arrives as ambiguous evidence. The system maintains a probabilistic model of what the user meant to write, commits to an interpretation, and renders the result in the document. The user keeps typing. There is no turn boundary, no discrete suggestion moment, no accept-or-reject gesture in the standard generative sense. ShapeWriter interprets gesture shapes. Handwriting recognizers interpret strokes ~\cite{zhaiShorthandWriting2003}. Autocorrect interprets noisy keypresses. Chinese pinyin IMEs interpret romanized phonetic sequences~\cite{huangMoonIME2018}. Swipe keyboards interpret continuous touch paths. All of these systems take initiative without being asked, on every input event, and what they take initiative on is interpretation rather than generation.

This second paradigm fits the workshop's working definition of proactivity (systems that anticipate user needs and act without explicit user input) as well as the first paradigm does. The taxonomy proposed around intelligence, adaptivity, and civility was designed for conversational systems with discrete turns, and several dimensions have no obvious analog in keystroke-level interpretation~\cite{dengHumancenteredProactive2024}. 
\citet{shiHCICentricSurvey2024} makes the scope explicit by limiting their survey to systems whose contribution is generated content. The result is that an entire family of proactive systems mostly sits outside the conceptual conversation. 

We propose to consider this family a distinct paradigm. %
\emph{Generative proactivity} covers systems whose initiative produces new content or action on the user's behalf: recommendations, suggestions, completions, appointments, drafts. 
\emph{Interpretive proactivity} covers systems whose initiative infers user intent from input the user has already provided. The most familiar cases are in text entry, where IMEs, autocorrect, and swipe keyboards each commit to reading of an ambiguous keystroke stream. The paradigm extends well beyond text as well. Gesture and handwriting recognition settle on what a movement or stroke was meant to be, and gaze-based interfaces infer which target the user intended. In every case, the system commits to one interpretation of input that the user has already produced.
Both take initiative under uncertainty, both inherit the decision-theoretic foundation Horvitz laid out, and both can be designed well or poorly~\cite{horvitzPrinciplesMixedinitiative1999}. Yet they differ on several axes that have practical consequences for how the design space should be charted.

\begin{figure*}[t]
    \centering
    \includegraphics[width=\linewidth]{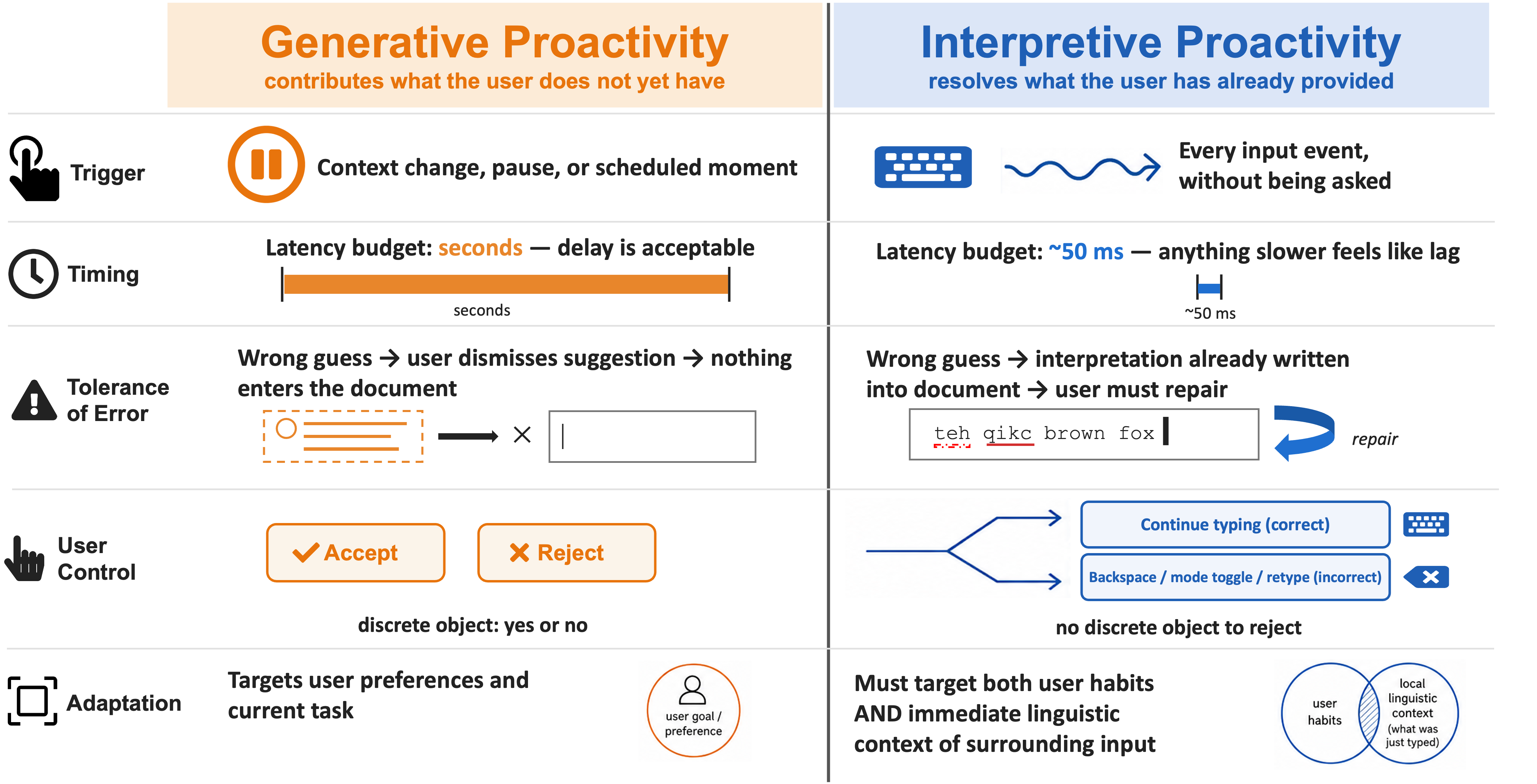}
    \caption{Comparison of Generative Proactivity and Interpretive Proactivity. Generative proactivity surfaces new artifacts at chosen moments; interpretive proactivity commits readings of existing input on every keystroke. The two paradigms differ in when they trigger, how much latency they can afford, what a wrong guess costs, how users regain control, and what adaptation must target.}
    \label{fig:comparison}
  \end{figure*}

The differences are clearest side by side (Figure \ref{fig:comparison}). 
Generative proactive systems are usually triggered by a context change, a pause, or a scheduled moment, while interpretive ones are triggered by every input event, often within milliseconds.  
The latency budget for the first is on the order of seconds; for the second, it is on the order of tens of milliseconds, because anything slower is felt as lag in the typing loop. 
When a generative system is wrong, the user dismisses a suggestion. When an interpretive system is wrong, the wrong interpretation has already been written into the document and has to be repaired. 
The unit of user control in the generative case is acceptance or rejection of a discrete artifact. In the interpretive case, it is something less standardized, ranging from backspace to mode toggles to phrase-level repair. 
Adaptation in the generative case usually targets the user's preferences or task. In the interpretive case, it has to target both the user's habits and the immediate linguistic context of the surrounding input.

We do not claim that interpretive proactivity needs an entirely new theory of its own.
The decision-theoretic core Horvitz identified, choosing among act, ask, and wait under uncertainty, carries over cleanly~\cite{horvitzPrinciplesMixedinitiative1999}. What does not carry over is the assumption that the act-ask-wait choice happens at the pace of user reflection. In interpretive proactivity, that choice happens thousands of times per session, and the design implications change. The rest of this paper works through those implications using two concrete cases.

\section{A Case Study: Multilingual Input and Auto-correction}

Two everyday systems show what interpretive proactivity looks like in practice: multilingual input methods and auto-correction. Both run on every keystroke, deciding what the user meant and committing to that reading without being asked. 
 
Consider a writer who uses two languages in a single document, which is common in technical and academic writing for bilingual writers. The keyboard has to decide, at each moment, which language the current keystrokes belong to. Most current systems push this decision back onto the user through a manual mode toggle: the writer presses a key to switch the input from one language to another, then presses it again to switch back. A short switch of a single word is easy to handle this way. A long stretch of mixed text, with technical terms in one language sitting inside sentences in another, turns the toggle into a steady source of interruption. The decoders for each language already work well on their own. What is missing is a way for the system to decide which decoder applies to which span without making the user state it by hand.
 
Auto-correction faces a structurally similar decision. A noisy or misspelled sequence of keypresses can map to several possible words, and the system has to choose one and commit it, often before the user has finished looking at the screen. When it chooses well, typing feels effortless. When it chooses badly, the user has to notice the wrong word, move the cursor back, delete it, and retype, sometimes fighting the same wrong correction a second time. The interpretation is silent and continuous, and the only signal the user gives is to keep typing or to go back and repair.

These two cases share the features that make interpretive proactivity distinct. The system acts on every input event, not at a moment of its choosing. It commits to one reading of the input instead of laying out candidates for the user to weigh. And the cost of a wrong commit is paid in repair work that interrupts writing, rather than in a dismissed suggestion. They also differ in an instructive way. Auto-correction usually has a reasonable recovery path, since undoing one word is quick. Multilingual input can be much more expensive, because a wrong language commit may disrupt a longer span of work and force a mode switch on top of the retype. This difference in repair cost is one reason the two systems, despite their shared structure, may call for different design choices.
We have been building an interactive prototype that varies systems like these along several design dimensions, which we treat as a design probe and report separately. 

\section{What Interpretive Proactivity Forces Us to Rethink}

\textbf{Timing.} In a generative system, timing means choosing when to interrupt the user with a suggestion. \citet{kuangEnhancingUX2024} studied exactly this and found that suggestions were received best when they arrived after the user had formed their own view, as a confirmation rather than a warning. In an interpretive system, timing means something different. It is the choice of when to commit an interpretation that the user never explicitly asked for. Committing early keeps typing fluid, but risks writing the wrong thing into the document. Committing late lets more context arrive and improves the guess, but the delay shows up as lag and breaks the feel of typing. The timing question here is not when to speak up. It is how long to wait before locking in a decision, weighed against how much delay the user will accept.
 
\textbf{Transparency.} In a generative system, transparency usually means showing the user why a suggestion was made, through a confidence score, a short explanation, or a highlighted reason. This works because suggestions are occasional, so there is room on the screen and in the user's attention to explain them. An interpretive system decides on every keystroke, so the same approach does not fit. Explaining every interpretation would flood the screen and pull the user out of writing. Even ordinary word suggestions carry a real attention cost, enough to slow people down despite the keystrokes they save~\cite{quinnCostBenefitStudy2016}. There is a further trap. When a system marks its low-confidence guesses, users catch the marked mistakes more often but catch the unmarked ones less, because their attention follows the marks, so the net gain can be close to zero~\cite{kristenssonFiveChallenges2009}. The candidate list in an input method is the narrow compromise that interpretive systems have settled on. It shows a few options without explaining any of them. The open question is which interpretations are worth surfacing at all, since the user cannot afford to look at most of them.
 
\textbf{Control.} In a generative system, control means accepting or rejecting a suggestion. The suggestion is a discrete object, so a single yes-or-no action is enough. In an interpretive system, the thing the user wants to fix is not a discrete object. When the system has read a whole phrase the wrong way, there is no suggestion sitting there to reject. The user is left with low-level tools, usually backspace, to wipe out a chunk of committed text and redo it by hand. What is missing is a control action at the right level: a way to reinterpret a span of text, such as flipping it to the other language or to a different correction, without deleting and retyping it. Previous research found that writers, especially those working in a second language, interact with suggestions in more varied ways than a plain accept-or-reject~\cite{buschekImpactMultiple2021}, which points to the same gap. Naming reinterpretation as a control action in its own right, alongside accept and reject, is one concrete thing the proactivity literature could take from the interpretive case.
 
\textbf{The cost of errors.} In a generative system, a wrong guess is inconsequential. The system offers a suggestion, the user ignores it, and nothing is lost. The mistake never enters the document, so the cost of being wrong is close to the cost of glancing at a suggestion and moving on. In an interpretive system, a wrong guess is costly because the system has already committed it. The mistaken interpretation is now sitting in the text, and the user has to notice it, go back, remove it, and redo the work, sometimes against a system that makes the same wrong guess again. 
 
The size of that cost is not fixed, and the difference between our two cases makes the point. When auto-correction picks the wrong word, the repair is small: undo one word and type it again. When a multilingual input method commits the wrong language across a longer span, the repair is much larger, because the user may lose a whole stretch of work and have to switch modes on top of retyping it. Two systems with the same underlying structure, deciding what input means on every keystroke, can therefore call for different designs, simply because being wrong costs so much more in one than in the other. A system whose mistakes are expensive should lean toward waiting, showing its uncertainty, and giving the user easy ways to reinterpret. A system whose mistakes are inconsequential can afford to act more freely. The cost of errors, then, is not a side issue for interpretive proactivity. It is the variable that should drive most of the other design choices.

\section{Open Questions for the Workshop}
\textbf{How should we measure the cost of a wrong interpretation?} Speed and keystroke savings are the usual metrics for text entry, but neither captures the effort of noticing a wrong commit and repairing it. Previous research has found that faster input is not always what people prefer~\cite{quinnCostBenefitStudy2016, buschekImpactMultiple2021}, which suggests that evaluating interpretive systems means weighing repair effort and interruption alongside raw efficiency.

\textbf{What mental model do users form of a system that interprets rather than suggests?} A visible suggestion is easy to read as the work of a separate agent, while a silent interpretation may feel like part of the keyboard itself. The difference likely shapes how much users trust the system and how they react when it is wrong.

\textbf{Can one system be interpretive and generative at the same time, and what happens when it is?} The two modes are not mutually exclusive, and in fact interpretation often underlies generation: a system must interpret what the user has written before it can generate a useful suggestion. 
Our claim is not that systems divide cleanly into interpretive and generative ones, but that interpretive initiative is a distinct mode that has gone underexamined even though it shapes design in its own way. The open question is what happens when one interface foregrounds both at once. 
Many input tools already blend the two, interpreting keystrokes while also offering whole-phrase suggestions. Combining both kinds of initiative in one interface may raise tensions that neither raises alone, and existing guidelines for AI-infused interfaces~\cite{amershiGuidelinesHumanAI2019} were written mainly for the generative case. 

\bibliographystyle{ACM-Reference-Format}
\bibliography{sample-base}
\end{document}